\documentclass[graybox]{svmult}

\usepackage{amssymb}
\usepackage{latexsym}
\usepackage{amsmath}
\usepackage{color}
\let\normalcolor\relax

\usepackage{mathptmx}       
\usepackage{helvet}         
\usepackage{courier}        
\usepackage{type1cm}        
\usepackage{makeidx}         
\usepackage{graphicx}        
\usepackage{multicol}        
\usepackage[bottom]{footmisc}

\usepackage{epsfig}
\usepackage[hang]{subfigure}
\usepackage{epstopdf}

\newcommand{\be}{\begin{eqnarray}}
\newcommand{\ee}{\end{eqnarray}}

\makeindex             

\begin{document}

\title*{Critical points in Lovelock Black Holes}
\author{Antonia M. Frassino, Robert B. Mann and Fil Simovic}
\institute{Antonia M. Frassino \at Frankfurt Institute for Advanced Studies, Ruth-Moufang-Stra\ss e 1, 
D-60438 Frankfurt am Main, Germany, \email{frassino@fias.uni-frankfurt.de}
\and Robert B. Mann \at Department of Physics and Astronomy, University of Waterloo,  Ontario N2L 3G1, Canada \at Perimeter Institute for Theoretical Physics, 31 Caroline St. N.,Waterloo, Ontario N2L 2Y5, Canada 
\and Fil Simovic \at Department of Physics and Astronomy, University of Waterloo, Ontario N2L 3G1, Canada
}
%
%
\maketitle
\abstract{
We review some of the results obtained by introducing a thermodynamic pressure via the cosmological constant in a class of higher curvature theories known as Lovelock gravity. In particular, we focus on a specific relation between the higher-order Lovelock couplings that introduces a peculiar isolated critical point for hyperbolic black holes characterized by non-standard critical exponents.
}

\section{Introduction}

For several  decades, the study of black hole (BH) thermodynamics has provided crucial information about the underlying structure of the spacetime. In particular, the fact that the BH entropy is proportional to the BH surface area in Planck units seems to tell us something important regarding the microstructure of spacetime.

Recently, a new development in BH thermodynamics based on scaling arguments and the Smarr relation \cite{Smarr:1972kt} was suggested  \cite{Kastor:2009wy} and led to the proposal that the mass of a BH in asymptotically anti de-Sitter (AdS) background should be interpreted as the enthalpy $H$ of the spacetime instead of the internal energy $U$. 
This total \textit{gravitational enthalpy} of the system $H=U+PV$ takes into account the possibility of representing the cosmological constant $\Lambda$ as a pressure $\Lambda = - P/(8 \pi G_N ) $ and define the thermodynamically conjugate variable to be the thermodynamic volume. Consequently, this proposal suggests to include in the first law of BH thermodynamics the variation of physical ``constants"  \cite{Kastor:2009wy,PhysRevD.52.4569,Cvetic:2010jb}
\begin{equation}
    dM = TdS + PdV
\end{equation}
where the thermodynamic BH volume is, by definition,
$    V= \left.\frac{\partial H}{\partial P} \right|_{S}\,,$
and can be different from the geometrical volume.

The \textit{extended thermodynamic phase space} assumed by this proposal is well motivated for at least three reasons:
\begin{enumerate}
\item In this extended phase space both the Smarr relation and the first law of thermodynamics hold, while in the conventional phase space only the first law is satisfied for nonzero  $\Lambda$. 

\item The use of an extended thermodynamic phase space is compatible with considering more fundamental theories of physics that admit the variation of physical constants.

\item Comparing the physics of BHs with real world thermodynamic systems becomes a much more reasonable possibility \cite{Kubiznak:2014zwa}.
\end{enumerate}

This proposal has been shown to provide a rich variety of thermodynamic behaviors for both AdS and dS black holes \cite{Kubiznak:2012wp, Kubiznak:2015bya}.  For example, by introducing the pressure, it can be shown \cite{Kubiznak:2012wp} that charged BHs behave as Van der Waals fluids\footnote{Van der Waals behavior is also present in the case of a regular BH in \cite{Frassino:2015hpa}.}.
It has been found that there could be triple points (e.g., in Kerr-AdS black holes \cite{Altamirano:2013uqa, Altamirano:2014tva}), where a coalescence of small, medium, and large sized BHs merge into a single kind at a particular critical value of pressure and temperature. Also, reentrant phase transitions \cite{Altamirano:2013ane} can occur, in which there are phase transitions from large BHs to small ones and then back to large again as the temperature monotonically increases.
In addition, this proposal has been subject to the attempts of study an extended AdS/CFT dictionary (see \cite{CJ, BPD, KRT, KR} and for an extended review \cite{KMT}).
 
\section{Lovelock Gravity}
Lovelock theory is a particular massless metric theory of gravity in arbitrary dimensions that, in four dimensions, become identical to general relativity with cosmological constant $\Lambda$. One of the main features of this model is that,  although the action functional of the theory could be an arbitrary higher-order polynomial  in the Riemann tensor, it leads to equations of motion that contain only second order derivatives of the metric tensor. 

To define the generic Lovelock densities one can use the language of differential forms or the generalized Kronecker delta symbols, where $\delta_{B_{1} \ldots B_{n}}^{A_{1}\ldots A_{n}}=n! \delta_{\left[B_{1}\ldots B_{n}\right]}^{A_{1}\ldots A_{n}}$
 is antisymmetric in any pair of upper and lower indices. In fact, one has that
$\delta_{B_{1} \ldots B_{n}}^{A_{1}\ldots A_{n}}$ is equal to $\epsilon_{B_{1}\ldots B_{n}}\epsilon^{A_{1}\ldots A_{n}}$
with respect to the Levi-Civita symbol. Using this definition, the Lovelock densities can be written as the complete contraction of the above generalized delta symbol with the Riemann curvature tensor:
\begin{equation}
\mathcal{L}_{n}=\frac{1}{2^{n}}\epsilon_{B_{1}\ldots B_{n}}\epsilon^{A_{1}\ldots A_{n}}\,R_{A_{1}A_{2}}^{\quad B_{1}B_{2}}\,\ldots\,R_{A_{2n-1}A_{2n}}^{\quad \: B_{2n-1}B_{2n}}. 
\label{eq:LLd}
\end{equation}
Using Eq.\eqref{eq:LLd}, one can check that the lowest order term $\mathcal{L}_{0}$  corresponds to a cosmological constant, while $\mathcal{L}_{1}$ is the Einstein-Hilbert term
\begin{eqnarray}
\mathcal{L}_{1} &=& \frac{1}{4}\left(2! \: \delta^{A_1 A_2}_{\left[ B_1 B_2 \right]} R^{\quad \: \: B_1 B_2}_{A_1 A_2}\right) = \frac{1}{2} R \label{L1}
\end{eqnarray}
and $\mathcal{L}_{2}$ is the Gauss-Bonnet combination:
\begin{eqnarray}
\mathcal{L}_{2} &=& \frac{1}{2^{4}}\left(4! \: \delta^{A_1 A_2 A_3 A_4}_{\left[ B_1 B_2 B_3 B_4 \right]} R^{\quad \: \: B_1 B_2}_{A_1 A_2} R^{\quad \: \: B_3 B_4}_{A_3 A_4}\right)\,.
\end{eqnarray}

\begin{figure*}
\centering
\begin{tabular}{cc}
\includegraphics[width=0.44\textwidth,height=0.28\textheight]{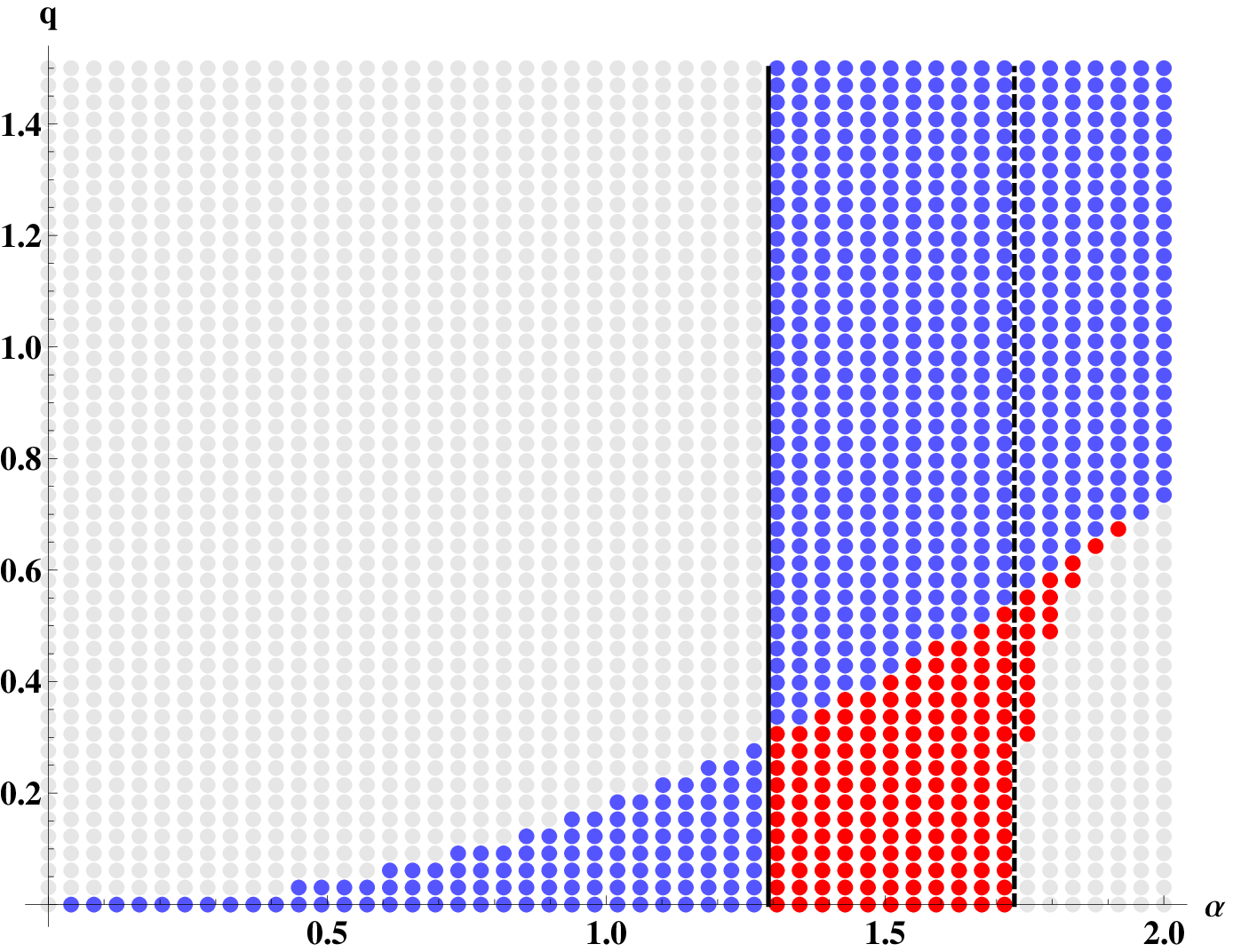} &
\rotatebox{0}{
\includegraphics[width=0.44\textwidth,height=0.28\textheight]{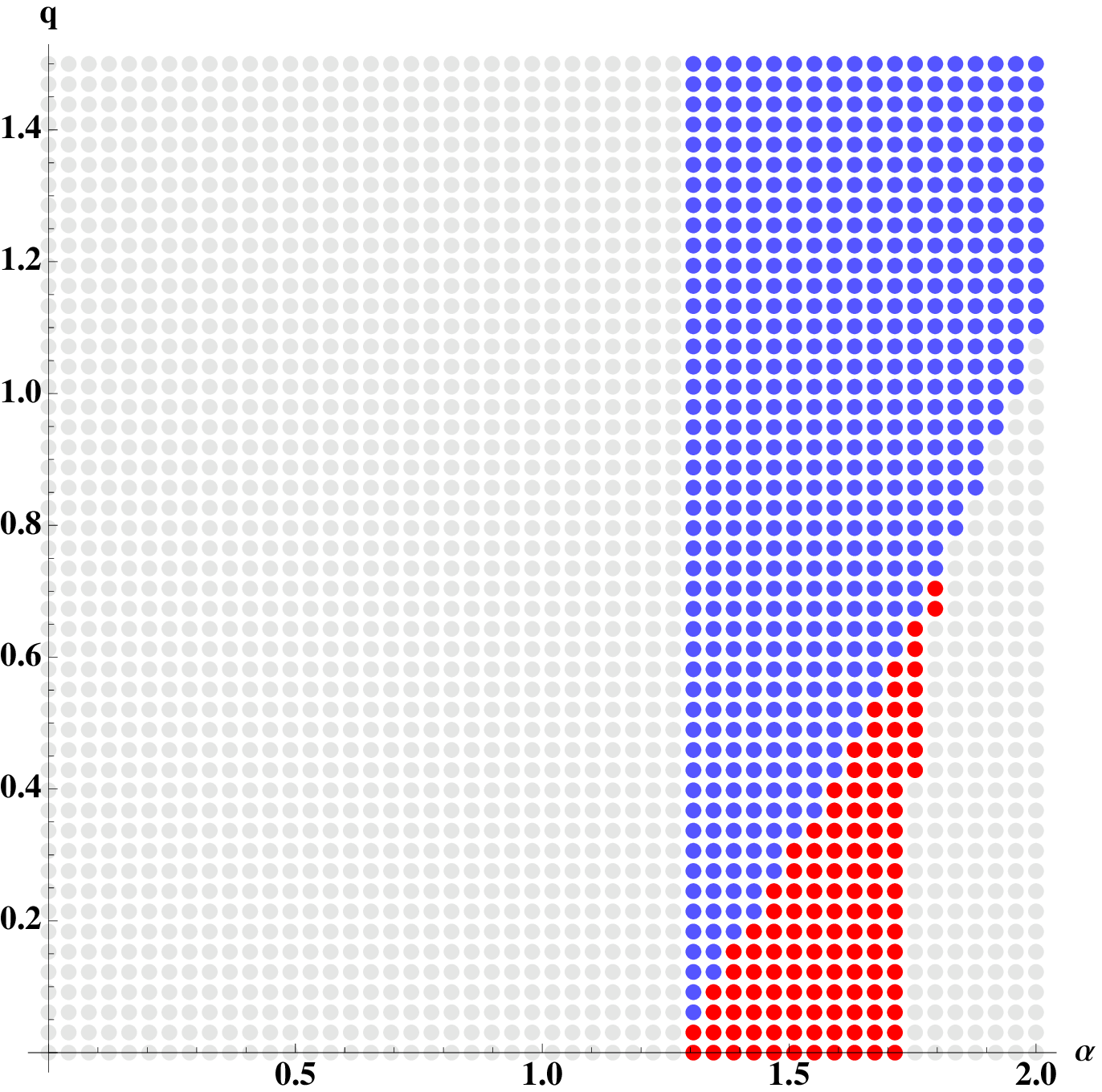}}\\
\end{tabular}
\caption{{\bf Critical points in $(q,\alpha)$-parameter space: $d=7$ ({\em left}) and $d=8$ ({\em right}), $\kappa=-1$ case.}
The figure shows the number of critical points with positive  $(P_c, V_c, T_c)$ and the opportune normalization (see details in \cite{Frassino:2014pha}) in the $(q,\alpha)$-parameter space. Grey dots correspond to no critical points, blue to one critical point, and red to two; black solid and dashed lines highlight $\alpha=\sqrt{5/3}$ and $\alpha=\sqrt{3}$, respectively.
Contrary to $d=7$ ({\em left}) case, in $d=8$ ({\em right}) there are no critical points for $\alpha<\sqrt{5/3}$.
}  
\label{Fig:Lvq7a}
\end{figure*}

\begin{figure*}
\centering
\begin{tabular}{cc}
\includegraphics[width=0.44\textwidth,height=0.28\textheight]{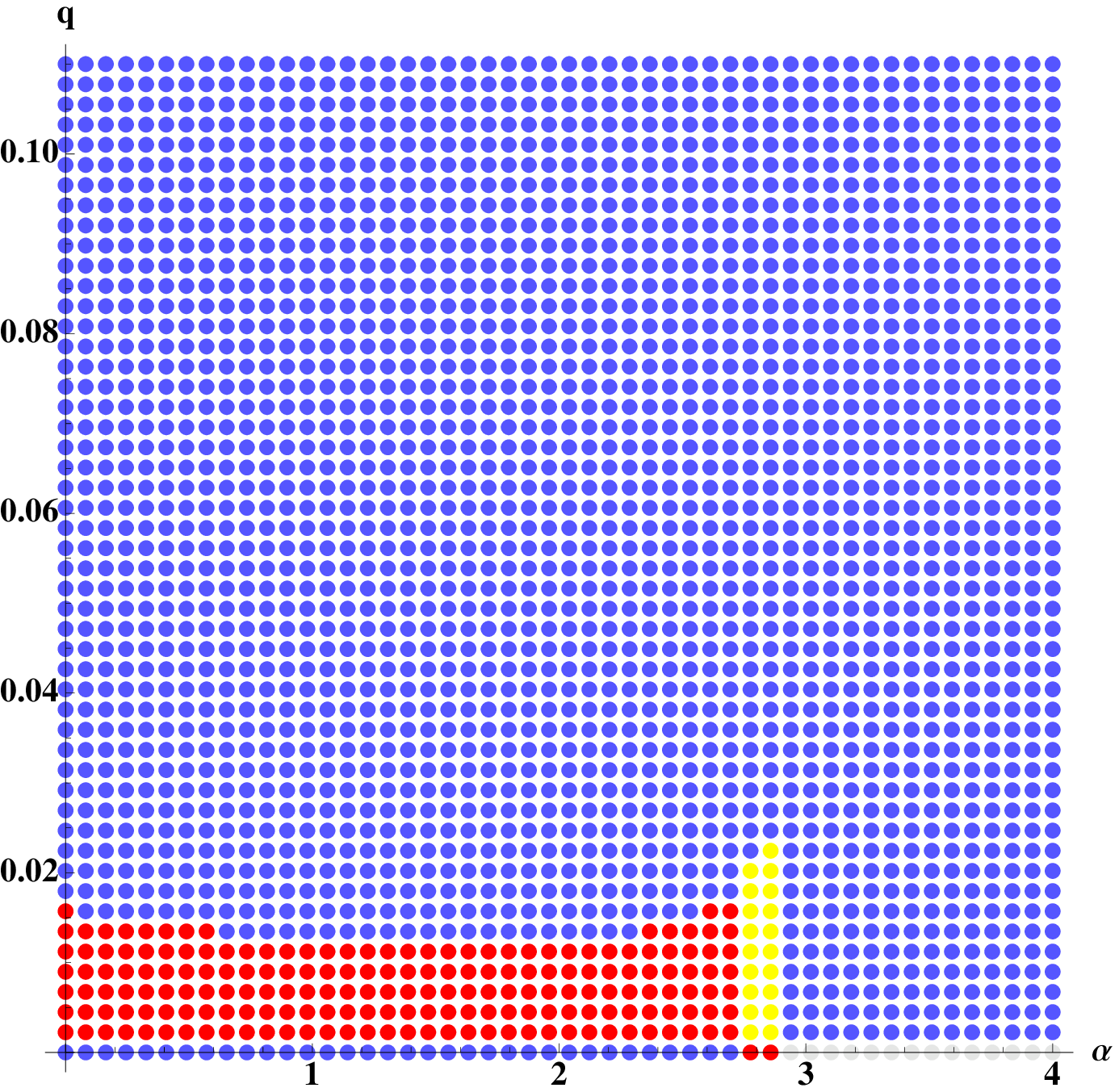} &
\rotatebox{0}{
\includegraphics[width=0.44\textwidth,height=0.28\textheight]{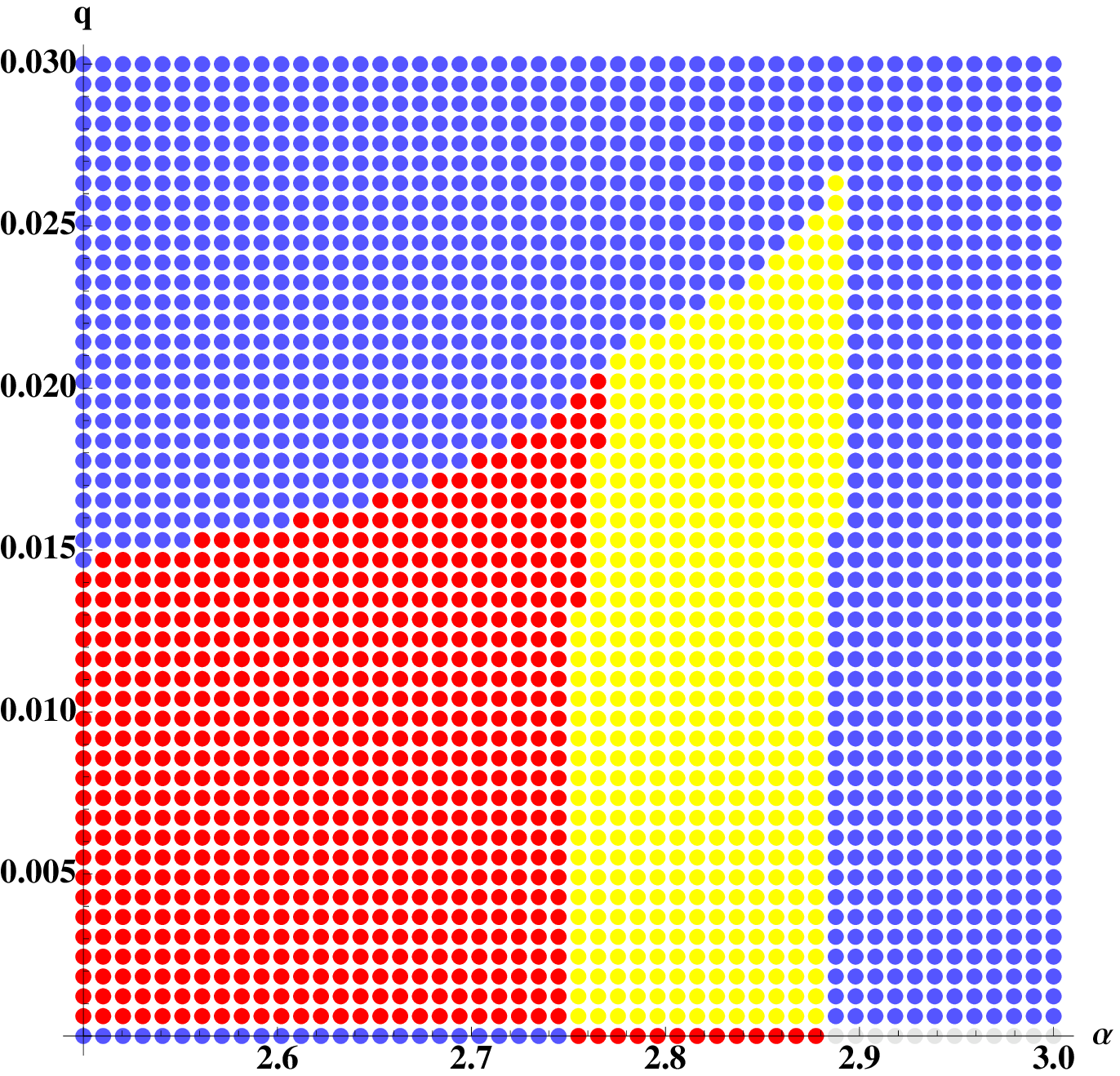}}\\
\end{tabular}
\caption{{\bf Critical points in $(q,\alpha)$-parameter space: $d=8, \kappa=+1$ case.} 
The number of critical points with positive  $(P_c, V_c, T_c)$ and the opportune normalization (see details in \cite{Frassino:2014pha}) is displayed in the $(q,\alpha)$-parameter space; grey dots correspond to no critical points, blue to one critical point, red to two, and yellow to three. The corresponding diagram for $d=7$ is trivial (contains only the blue region with one critical point) and hence is not displayed. Although all critical points have positive  $(P_c, V_c, T_c)$, some $P_c$ may exceed the maximum pressure $p_+$ and hence occurs for a compact space.  Note also the qualitatively different behavior for $q=0$.
}  
\label{Fig:Lvq7b}
\end{figure*}

This means that the combinations of all the Lovelock densities for $n>D$ are zero, whereas when $n \equiv D$ (the \textit{critical dimension} for a given Lovelock term), the Lovelock density is topological: 
it is a total derivative making no contribution to the field equations and whose value depends on the topology of the spacetime manifold.  The simplest example is the Einstein-Hilbert term in two dimensions:  the Ricci scalar is a total derivative and the Einstein tensor is identically zero\footnote{The Gauss-Bonnet theorem relates the topological ``Euler number'' $\chi$ of a two dimensional surface $\mathcal{M}_2$ defined as $ \chi \left[\mathcal{M}_2 \right]=2-2h$ where $h$ is the number of topological handles to a differentiable geometric quantity, the scalar curvature, in this way:\begin{equation}
\chi \left[\mathcal{M}_2 \right]=\frac{1}{4 \pi} \intop_{\mathcal{M}} R.
\end{equation} Chern generalized the theorem to higher dimensions finding the relevant higher-order curvature scalars. Thus the Gauss-Bonnet density, for example, is a topological invariant in four dimensions whose integral is the generalized Euler or Chern topological number.  }.
Thus, the Lovelock densities are a generalization of the Chern scalar densities, namely densities whose variation leads to a second order equation of motion. Any higher order derivatives present in the variation of Lovelock densities end up as total divergences and thus do not contribute to the field equations. For example, the six-dimensional Euler density in seven or eight dimensions (used in Figs.\ref{Fig:Lvq7a} and \ref{Fig:Lvq7b}) will be a Lovelock density of third power in the curvature tensor (we refer to it as 3rd-order Lovelock gravity).
 
The generic Lovelock Lagrangian is given by 
\begin{equation}
\label{eq:Lagrangian}
L=\sum_{n=0}^{k_{max}} \hat{\alpha}_{\left(n\right)} \mathcal{L}_{n}
\end{equation} 
where $k_{max}$ is the integer part of $\left[ \left( D-1 \right)/2 \right]$ and the $\alpha$'s are the Lovelock coupling constants.
In the \emph{extended thermodynamic phase
space}, the Lovelock coupling constants (including the cosmological
constant $\hat{\alpha}_{\left(0\right)}$) can be considered as thermodynamic variables and can vary in the first law of BH thermodynamics.
The physical meaning of these variables and their conjugates, apart from the cosmological constant and its associated volume,
remains to be explored.\footnote{A similar situation was seen in Born--Infeld electrodynamics,
in which the thermodynamics conjugate to the Born--Infeld coupling constant was interpreted as vacuum polarization \cite{Gunasekaran:2012dq}. }

\subsection{Thermodynamic Considerations}\label{sec:LCBH}

Let us consider now a Lovelock BH charged under a Maxwell field, $F=dA$, with the action given by (cf. Eq.\eqref{eq:Lagrangian}) 
\begin{equation}
I=\frac{1}{16\pi G_{N}}\int d^{d}x\sqrt{-g}\Bigl({\sum_{k=0}^{k_{max}}}\hat{\alpha}_{\left(k\right)}\mathcal{L}_{k}-4\pi G_{N}F_{ab}F^{ab}\Bigr)\,,\label{eq:Loveaction}
\end{equation}
and the corresponding equations of motion 
\begin{equation}
\sum_{k=0}^{k_{max}}\hat{\alpha}_{\left(k\right)}\mathcal{G}_{ab}^{\left(k\right)}=8\pi G_{N}\Bigl(F_{ac}{F_{b}{}^{\,c}}-\frac{1}{4}g_{ab}F_{cd}F^{cd}\Bigr)\,\label{eq:Graveq}
\end{equation}
where $\mathcal{G}_{ab}^{\left(k\right)}$ is the generalized Einstein's tensor.
The Hamiltonian formalism admits a derivation of the expression for the gravitational entropy and the corresponding first law of BH thermodynamics \cite{Jacobson:1993xs}. In \cite{Kastor:2010gq}, both the first law and the associated Smarr formula in an extended phase space can be obtained exploiting the Killing potential formalism. 
For a Lovelock BH, characterized by the mass $M$, the charge $Q$, the temperature $T$ and the entropy $S$, the extended first law and the associated
Smarr relation read \cite{Jacobson:1993xs,Kastor:2010gq} 
\begin{eqnarray}
\delta M & = & T\delta S-\frac{1}{16\pi G_{N}}\sum_{k}\hat{\Psi}^{\left(k\right)}\delta\hat{\alpha}_{\left(k\right)}+\Phi\delta Q\,,\label{first}\\
\left(d-3\right)M & = & \left(d-2\right)TS+\sum_{k}2\left(k-1\right)\frac{\hat{\Psi}^{\left(k\right)}\hat{\alpha}_{\left(k\right)}}{16\pi G_{N}}+\left(d-3\right)\Phi Q\,.\label{Smarr}
\end{eqnarray}
The potentials
$\hat{\Psi}^{\left(k\right)}$ are the thermodynamic conjugates to the $\hat{\alpha}_{(k)}$'s and are a non-trivial functions of the ``bare'' cosmological constant
$\Lambda=-\hat{\alpha}_{0}/2$ and of the higher-order Lovelock couplings \cite{Frassino:2014pha}. In general, in Lovelock gravity, the BH entropy is no longer given by one-quarter
of the horizon area, but rather reads 
\begin{equation}
S=\frac{1}{4G_{N}}\sum_{k}\hat{\alpha}_{k}{\cal A}^{(k)}\,,\quad{\cal A}^{(k)}=k\int_{\mathcal{H}}\sqrt{\sigma}{\mathcal{L}}_{k-1}\, \label{S}
\end{equation}
where $\sigma$ denotes the determinant of $\sigma_{ab}$, the induced
metric on the BH horizon ${\mathcal{H}}$, and the Lovelock
terms ${\mathcal{L}}_{k-1}$ are evaluated on that surface.
A curious feature is that the Lovelock black brane entropy (and also the other thermodynamic expressions when considered under an appropriate rescaling) does not depend on the Lovelock coupling constant $\hat{\alpha}_{k}$ for $k \geq 2$ (see \cite{Cadoni:2016hhd}).

In what follows, the (negative) cosmological constant $\Lambda=-\hat{\alpha}_{0}/2$
with the thermodynamic pressure and the conjugate quantity $\hat{\Psi}^{\left(0\right)}$
with the thermodynamic volume $V$, are identified in the following way
\begin{eqnarray}
P &=&-\frac{\Lambda}{8\pi G_{N}}=\frac{\hat{\alpha}_{0}}{16\pi G_{N}}\,,\label{P}\\
V &=&-\hat{\Psi}^{(0)}=\frac{16\pi G_{N}\Psi^{(0)}}{(d-1)(d-2)}=\frac{\Sigma_{d-2}^{(\kappa)}r_{+}^{d-1}}{d-1}\,,\label{V}
\end{eqnarray}
where $\Sigma_{d-2}^{(\kappa)}$ denotes the finite volume of the $(d-2)$-dimensional compact space 
at constant $(r,t)$, whose
constant curvature $(d-2)(d-3)\kappa$, with the horizon geometry corresponding to $\kappa = 0, +1,-1$ for flat (brane), spherical and hyperbolic black hole horizon geometries respectively. 
This identification allows interpreting the mass $M$ of the BH as an enthalpy rather than the internal energy of the system.
Using Eq. \eqref{P} and the thermodynamic volume given by Eq. \eqref{V} in the definition of the Hawking temperature, one can obtain the Lovelock ``fluid equation of state''\footnote{In terms of the rescaled Lovelock coupling constants \begin{equation}
\alpha_{0}=\frac{\hat{\alpha}_{(0)}}{\left(d-1\right)\left(d-2\right)}\,,\quad{\alpha}_{1}={\hat{\alpha}}_{(1)}\,,\quad\alpha_{k}=\hat{\alpha}_{(k)}\prod_{n=3}^{2k}\left(d-n\right){\quad\mbox{for}\quad k\geq2}\,.
\end{equation}} 
\begin{eqnarray}\label{eq:eqofstateL}
P & = & P(V,T,Q,\alpha_{1},\dots,\alpha_{k_{max}})= \nonumber \\
 & = & \frac{d-2}{16\pi G_{N}}\sum_{k=1}^{k_{max}}\frac{\alpha_{k}}{r_{+}^{2}}\Bigl(\frac{\kappa}{r_{+}^{2}}\Bigr)^{k-1}\Bigl[4\pi kr_{+}T-\kappa(d-2k-1)\Bigr]+\frac{Q^{2}}{2\alpha_{1}r_{+}^{2(d-2)}}\,,
\label{state}\end{eqnarray}
and study the possible phase transitions based on the behavior of
the Gibbs free energy in the canonical ensemble
$G=M-TS=G(P,T,Q,\alpha_{1},\dots,\alpha_{k_{max}})\,$. 
The equilibrium thermodynamic state corresponds to the global
minimum of this quantity for fixed parameters $P,T,Q$ and $\alpha$'s. 
A critical point occurs when $P=P(V)$ has an inflection point, i.e., when
\begin{equation}
\label{cp}
\frac{\partial P}{\partial V}=0\,,\quad \frac{\partial^2 P}{\partial V^2}=0\,.
\end{equation}
Together with the equation of state \eqref{eq:eqofstateL}, the system \eqref{cp} determines the critical values 
$\{P_c, V_c, T_c\}$ as functions of $Q$ and $\kappa$. To find a critical point one has to solve the (higher-order polynomial) Eqs. \eqref{cp} for $T_c, V_c$ and insert the result  into the equation of state \eqref{eq:eqofstateL} to find $P_c$, subject to the restriction that  $P_c, V_c, T_c$ have positive values in order for the critical points to be physical.
As result of this study, we find that critical behaviour occurs in $d = 7, 8, 9, 10, 11$
dimensions, but not $d=6$ (Gauss-Bonnet case), though there is a cusp  for $\kappa=+1$ .  If $d = 7$, the critical point is associated with  Van der Waals behavior, whereas
in $d = 8, 9, 10, 11$ we observe a reentrant phase transition.

Figures \ref{Fig:Lvq7a} and \ref{Fig:Lvq7b}
show the results of a numerical analysis of 3rd-order Lovelock gravity in $d=7, 8$ with $\kappa= \pm 1$ in terms of the following dimensionless variables:
\begin{equation}
r_{+}=v\,\alpha_{3}^{\frac{1}{4}}\,,\quad T=\frac{t\alpha_{3}^{-\frac{1}{4}}}{d-2}\,,\quad m=\frac{16\pi M}{(d-2)\Sigma_{d-2}^{(\kappa)}\alpha_{3}^{\frac{d-3}{4}}}\,,\quad Q=\frac{q}{\sqrt{2}}\alpha_{3}^{\frac{d-3}{4}}\,.\label{dimLov}
\end{equation}

For $\kappa=-1$, a special case occurs when the parameter $\alpha= \alpha_2/\sqrt{\alpha_3}   = \sqrt{3}$. The system can be resolved analytically, and the solution exhibits a special isolated critical point. The Gibbs free energy displays two swallowtails, both emerging from this point, and the critical exponents (obtained
by series-expanding the equation of state near the critical point) are
\begin{equation}
\tilde{\alpha}=0,\,\,\,  \tilde{\beta}=1,\,\,\,     \tilde{\gamma}=2,\,\,\,    \tilde{\delta}=3.
\end{equation}
for any dimension $d\geq 7$. Three of these critical exponents are independent because of a violation of certain scaling relations, in contrast to  two independent exponents in  mean field theory. This isolated critical point can be understood as the merging of two
critical points, and we find the BH is massless ($M = 0$)  \cite{Mann1} in this limit.  By comparison, taking $\kappa=+1$
for the same value of  $\alpha_2$ yields
\begin{equation}
\tilde{\alpha}=0,\,\,\,  \tilde{\beta}=\frac{1}{2},\,\,\,  \tilde{\gamma}=1,\,\,\,    \tilde{\delta}=3,
\end{equation}
which are  the standard swallowtail mean field theory critical exponents.

\section{Conclusions}

3rd-order Lovelock gravity presents interesting and qualitatively new thermodynamic behaviour. 
In particular, 3rd-order uncharged Lovelock black holes with $\alpha=\sqrt{3}$ and $\kappa=-1$ are especially peculiar: in this interesting, distinctive case, we find that the equation of state has a non-standard expansion around a special critical point suggesting a violation of the scaling relations and non-standard critical exponents. This feature of Lovelock gravity has been further discussed in \cite{Frassino:2014pha, Dolan:2014vba} and was subsequently observed in quasi-topological gravity as well
\cite{Hennigar:2015esa}.

 In general, since the odd-order Lovelock theories (in any dimension in which they exist) always admit massless topological black holes \cite{Mann1,Mann2} as solutions, they will all exhibit the peculiar isolated critical point for an appropriate choice of coupling constants.


\end{document}